\documentclass[fleqn,usenatbib]{mnras}

\usepackage{newtxtext,newtxmath}

\usepackage[T1]{fontenc}

\DeclareRobustCommand{\VAN}[3]{#2}
\let\VANthebibliography\thebibliography
\def\thebibliography{\DeclareRobustCommand{\VAN}[3]{##3}\VANthebibliography}

\usepackage{graphicx}	
\usepackage{amsmath}	

\usepackage{multirow}

\usepackage{soul}
\usepackage[normalem]{ulem}
\usepackage[dvipsnames]{xcolor}

\title[Optical mass proxies from cluster membership]{Galaxy cluster optical mass proxies from probabilistic memberships} 

\author[Doubrawa et al.]{
Lia Doubrawa,$^{1}$\thanks{E-mail: lia.doubrawa@usp.br}
Eduardo S. Cypriano,$^{1}$
Alexis Finoguenov,$^{2}$
Paulo A. A. Lopes,$^{3}$
Matteo Maturi,$^{4,5}$
\newauthor Anthony H. Gonzalez,$^{6}$
Renato Dupke$^{7,8,9}$
\\
$^{1}$ Instituto de Astronomia, Geof\'{\i}sica e Ci\^encias Atmosf\'ericas, Universidade de São Paulo, Rua do Mat\~ao 1226, 05508-090 São Paulo, Brazil\\
$^{2}$ Department of Physics, University of Helsinki, P.O. Box 64, FI-00014 Helsinki, Finland\\
$^{3}$ Observat\'orio do Valongo, Universidade Federal do Rio de Janeiro, Ladeira do Pedro Ant\^onio 43, Rio de Janeiro RJ 20080-090, Brazil\\
$^{4}$ Zentrum f\"ur Astronomie, Universitat\"at Heidelberg, Philosophenweg 12, D-69120 Heidelberg, Germany\\
$^{5}$ Institute for Theoretical Physics, Philosophenweg 16, D-69120 Heidelberg, Germany\\
$^{6}$ Department of Astronomy, University of Florida, Gainesville, FL 32611-2055, USA \\
$^{7}$ Observatório Nacional, Rua General José Cristino, 77, São Cristóvão, 20921-400, Rio de Janeiro, RJ, Brazil \\
$^{8}$ Department of Astronomy, University of Michigan, 311 West Hall, 1085 South University Ave., Ann Arbor, USA\\
$^{9}$ Department of Physics and Astronomy, University of Alabama, Box 870324, Tuscaloosa, AL, USA\\}

\date{Accepted 2023 September 29. Received 2023 September 29; in original form 2023 April 4}

\pubyear{2023}

\begin{document}
\label{firstpage}
\maketitle

\begin{abstract}
Robust galaxy cluster mass estimates are fundamental for constraining cosmological parameters from counts. 
For this reason, it is essential to search for tracers that, independent of the cluster’s dynamical state, have a small intrinsic scatter and can be easily inferred from observations.
This work uses a simulated data set to focus on photometric properties and explores different optical mass proxies including richness, optical luminosity, and total stellar mass. 
We have developed a probabilistic membership assignment that makes minimal assumptions about the galaxy cluster properties, limited to a characteristic radius, velocity dispersion, and spatial distribution.
Applying the estimator to over $919$ galaxy clusters with $z_{phot}<0.45$ within a mass range of $10^{12.8}$ to $10^{15}$ M$_\odot$, we obtain robust richness estimates that deviate from the median true value (from simulations) by $-0.01 \pm 0.12$. The scatter in the mass-observable relations is $\sigma_{log_{10}(M|\mathcal{R})}=0.181 \pm 0.009$ dex for richness, $\sigma_{log_{10}(M|L_\lambda)}=0.151 \pm 0.007$ dex for optical luminosity, and $\sigma_{log_{10}(M|M_\lambda^*)}=0.097 \pm 0.005$ dex for stellar mass. We also discuss membership assignment, completeness and purity, and the consequences of small centre and redshift offsets.
We conclude that the application of our method for photometric surveys delivers competitive cluster mass proxies.
\end{abstract}



\section{Introduction} 

\subsection{The need for optical-based mass-proxies for galaxy clusters}

Wide-field cosmological imaging surveys such as KiDS \citep{deJong2013}, DES \citep{des2005}, HSC \citep{Miyazaki2012}, WISE \citep{Wise2010}, LSST \citep{Ivezic2019}, EUCLID \citep{Sartoris2016, Euclid2022} and others, provide (or are expected to provide) datasets containing billions of galaxies. These data can help us understand the large-scale structure of the Universe and its evolution.  
As galaxy clusters trace overdensity peaks in the matter distribution, their abundance as a function of mass and redshift is quite sensitive to the matter density of the Universe and the evolution of its clustering. Thus, galaxy clusters are known as powerful tools for constraining cosmological parameters \citep[e.g.][]{Carlberg1996, Reiprich2002, Voit2005, Vikhlinin2009, Allen2011, Weinberg2013, Pacaud2016, Costanzi2019, Chitham2020, Finoguenov2020}. 
This approach, however, depends on robust and precise mass estimates for all detected structures. Since this cannot be directly inferred from observations, we must rely on observable proxies for the halo masses. 

An ideal mass proxy should have a small intrinsic scatter, present a minor dependence on the cluster's dynamical state, and be easily accessed through observations.
Various independent mass proxies have been explored in recent decades to minimize uncertainties and systematic effects. Statistical uncertainties can go down to $\sim 0.2$-$0.3$ dex for gas mass, temperature, X-ray luminosity \citep{Ettori2013, Sereno2019}, Sunyaev-Zel'dovich effect signal \citep{Planck2015, Pratt2020}, stellar mass  \citep{Pereira2018} and cluster optical richness \citep{Lopes2009b, Andreon2015, Bellagamba2019}. Recently, \cite{Costanzi2019} and \cite{Chitham2020} successfully utilized optical richness to derive cosmological constraints in SDSS \citep{Aihara2011} and CODEX \citep{Finoguenov2020} surveys, respectively. Nevertheless, statistics of low-mass structures such as galaxy groups are still limited and require further investigation.
 
Imaging surveys that use narrow-band filters are particularly interesting for galaxy cluster studies. Those surveys can be seen as a midpoint between broad-band imaging and spectroscopic surveys in the sense of the precision and accuracy of galaxies' photometric redshifts (photo-z's). Examples includes J-PAS \citep{Benitez2014, Bonoli2021}, PAU \citep{Marti2014}, S-PLUS \citep{Mendes2019} and J-PLUS \citep{Cenarro2019}.
In terms of group and cluster detections, this transition affects the cluster finder algorithms in the sense that the methodologies are migrating from a colour-based approach to ones that can take advantage of the increasing spectro-photometric information \citep[for example, the performance and algorithm selection in][]{Euclid2019III}.   

It is, therefore, also necessary to revisit the membership estimate techniques and the strategies that use the spectro-photometric data to derive mass proxies with low scatter. However, a common problem when dealing with photometric redshifts is the magnitude of the uncertainties in the redshift space. A typical error of $0.01$ in photo-zs corresponds to $3000$ km/s, which is $3$-$6$ larger than a typical, virialized, cluster velocity dispersion estimated with spectroscopic redshifts. This effect leads to contamination by neighbouring galaxies along the line of sight and can also create false cluster detections by linking unbounded structures \citep{Weinberg2013}. The use of less discriminating observables, such as individual colours, can exacerbate this issue. 

Some recent works by \cite{Castignani2016, Bellagamba2019, Lopes2020} have addressed this matter regarding the membership of galaxies to clusters or groups. The use of continuous probabilistic memberships, instead of binomial ones (member/non-member), emerges as a superior description of what can be inferred given the data available. This approach allows for the definition of a set of galaxy-based mass proxies that can be derived by weighting galaxy numbers, luminosities, stellar masses, or others according to their membership probability.

\subsection{The state-of-the-art in probabilistic membership}

In 2000, \citeauthor{Brunner2000} introduced a technique that produces a probabilistic membership interpretation without directly relying on spectroscopic information. First, the authors assign a Gaussian probability distribution function to each galaxy centred on the estimated photometric redshift. Next, the membership is obtained by integrating the distribution function within the galaxy cluster redshift interval.

This method was improved by \cite{George2011} and \cite{Castignani2016}. The authors introduce a Bayesian formalism that also considers the cluster centre-galaxy projected distance, magnitudes, and the relative population size between field and cluster galaxy density. These studies, tailored for the specific sample in question, provide a better understanding of the membership dependence on photo-z accuracy, magnitudes, and cluster properties. 

In the context of previous studies of galaxy cluster detection and galaxy population identification, the redMaPPer \citep{Rykoff2014} and AMICO \citep{Bellagamba2019, Maturi2023} cluster-finders advanced meaningfully by relying on an optimal filtering formalism. Considering the previously commented information, the methods also included self-similar models to describe the expected magnitude and galaxy distributions in clusters. This modelling allowed better discrimination between member and field galaxies, producing membership estimates with lower contamination.

In this work, we aim to study probabilistic memberships and their possible application as photometric-based mass proxies by expanding previous approaches to explore well-defined photo-z probability density functions (PDFs) surveys like S-PLUS and J-PAS enable. The procedure allows further refinements over the red sequence methods and includes fainter galaxies, essential for galaxy groups. Here, we take a step back and revisit potential results that can be achieved without strong modelling hypotheses and possible priors that are not entirely adequate to our data.

Motivated by the advent of several ongoing and forthcoming wide photometric surveys, we introduce a galaxy membership algorithm that follows the minimal assumption, data-driven, and photo-z PDF-based approach. We analyze the galaxy group/cluster distributions down to $\log_{10}(M) > 12.8$ M$_\odot$, using a mock sky simulation based on the Southern Photometric Local Universe Survey \citep[S-PLUS, ][]{Mendes2019} up to redshift $z = 0.45$, and test several physically meaningful mass-proxies that can be derived from those memberships.
To the best of our knowledge, only \cite{Castignani2016}, \cite{Bellagamba2019}, and \cite{Werner2022} consider samples through this mass range.

\subsection{Outline of the paper}

In \S\ref{Data} we present details of the simulated sky area. We next describe in \S\ref{Richness_code} membership assignments based on a variable aperture radius. In \S\ref{Results} we show the results from applying the method over the mock catalogue, focusing on the performance of the richness estimation, scaling relations with different optical proxies, and minor disturbances in the cluster centre and redshift. In \S\ref{Individual Memberships} we discuss the membership significance, present completeness and purity results, and derive cluster positions and $z$ based on member galaxies.
Finally, in \S\ref{Conclusions} we summarize the results and present the conclusions.

Through this work, we adopt a flat $\Lambda$CDM cosmology, with $h=0.673$, $\Omega_m=0.315$ and $\Omega_\Lambda=0.685$, following the \cite{Planck2014} parameters, which are the same cosmology adopted in the simulations. Magnitudes are given in the AB system.

\section{Mock catalogue} \label{Data} %

To evaluate our estimators' performance, we use a mock catalogue from a simulated sky lightcone by \cite{Araya-Araya2021, Werner2022}, made to emulate the S-PLUS\footnote{S-PLUS is a photometric survey of the Southern Sky using 12 optical bands (7 narrow and 5 broad), from a 0.8m robotic telescope located at Chile/CTIO \citep{Mendes2019}.} survey first data release \citep{Mendes2019}. 

This mock catalogue simulates a projected area of 324 square degrees, using synthetic galaxies following \cite{Henriques2015} analytical model (SAM). The algorithm uses the Millenium run simulation \citep{Springel2005b} as a base to generate an equivalent matter density field scaled by the Planck 1 cosmological framework \citep{Planck2014}. The general halo mass resolution is $m_p=9.6\, 10^8$ M$_\odot/ h$, but only those with the corresponding stellar mass higher than $10^8$  M$_\odot/ h$ are accepted in the simulation.

All the dominant dark matter halos with $M_{c,200} \geq 10^{12.8}$ M$_\odot$ are selected. To ensure robust membership estimates and statistics, we choose only the ones with at least $3$ associated galaxies. 
The associated galaxy members (hereafter ``true members'') are retrieved from the merger history of the system. All galaxies that formally reside in a dark matter halo and evolve into a chosen cluster receive the cluster identification ID ``haloId''. This ID allows us to easily identify all the galaxies that belong to a certain cluster. The sky coordinates are estimated from the median value of the member galaxies' distribution. Limiting the catalogue to $z<0.45$, we obtain $238$ groups with masses between $10^{12.8}$--$10^{13.5}$ M$_\odot$, $358$ clusters with $10^{13.5}$--$10^{14}$ M$_\odot$, $249$ with $10^{14}$--$10^{14.5}$ M$_\odot$, and $76$ massive clusters with $M> 10^{14.5}$ M$_\odot$.
One limitation of the catalogue is that the statistics at low and high masses are poor due to the simulation mass limits. We cannot state how representative the sample is in relation to the full halo population and the impact on the results. A detailed analysis of the simulation is beyond the scope of this paper. Low masses are also affected by the applied membership cut (see the resulting distributions in Fig.\,\ref{fig:results}). 

Observed true redshifts in the mocks are created by adding cosmological redshifts (assuming that all galaxies at $z_i$ have a comoving distance of $d_C(z_i) < d_{C,gal}<d_C(z_i)+30$ kpc) and Doppler redshifts created by peculiar motions of each mock galaxy. Simulation 3D coordinates are generated following the \cite{Kitzbichler2007} methodology.

Star formation histories (SFH) and stellar masses are extracted from the output of the SAM. With the evolution of the information through the simulation's cosmic time, it is possible to attribute spectral energy distributions (SED) to each time step. SEDs, alongside dust extinction models, are used to compute u, g, r, i, and z apparent magnitudes, similarly to the Sloan AB filter system \citep{Fukugita1996}.

The mock photometric redshifts are generated by perturbing the true value by adding a random number from a normal distribution with $\sigma_{MAD,z}$ as the standard deviation. $\sigma_{MAD,z}$ is the normalized median absolute deviation of the comparison between photo-zs and spectroscopic redshifts for a given galaxy magnitude $r$ obtained for the DR1/S-PLUS survey \citep[$\sigma = MAD(r)$, ][with a median value of $\sigma=0.026$]{Molino2020}.

A similar procedure generates the galaxy's photo-z probability density function (PDF). For each galaxy, we assume a normal distribution with $\sigma = MAD(r)$ as the standard deviation, centred on the created photometric redshift.
This technique considers the correlation between the galaxy's magnitudes and its photometric redshift errors.

\section{A density-based probabilistic membership algorithm} \label{Richness_code}

\subsection{A fixed aperture richness estimator (FAE)}

To make a first rough richness estimation, we calculate the galaxy overdensity at a given position in the projected (celestial) coordinates and redshift, or the (2+1)D space for a given cluster.

The galaxies are selected based on the distance from the cluster centre, within a cutoff radius of $500$ kpc \footnote{as the median value of the characteristic radius $R_c$ of the cluster sample, more details are provided in \S\,\ref{sec:AME} and \ref{sec:rc}}, and a redshift interval, $z_{cluster} \pm \Delta z$ (with $\Delta z = 3\sigma_{MAD,z}[1+z_{cl}]$), where $\sigma_{MAD,z}[1+z_{cl}]$ is a robust estimation of the photo-z scatter at the relevant redshift \citep[similar to][]{Araya-Araya2021}. 

A probability membership can then be assigned to each of the selected galaxies by:
\begin{equation} \label{eq:cdf}
    P_{clz} = \int_{z_{cl}-\Delta z}^{z_{cl}+\Delta z} z_{phot}(z^\prime) dz^\prime
\end{equation}
\noindent where $z_{phot}$ is the photo-z PDF.

The richness around the cluster position is estimated as, 
\begin{equation}
   \mathcal{R} = \sum_i{P_{clz,i}} - \Sigma_{fld}(z_{cl}) \times A_{cl} 
   \label{eq:rich_fix}
\end{equation}
where i is the index that runs over all galaxies within the cluster area ($A_{cl}$), and $\Sigma_{fld}(z)$ is the median surface density of field galaxies at a given redshift. We calculate the surface density as the sum of the galaxy's probabilities per area for randomly distributed sky coordinates. We avoid bias due to the presence of clusters or voids with a $3 \sigma$ clipping procedure.
This strategy is adequate for small/medium sky areas.

Another possible approach to dealing with field galaxies contamination is to compute the contribution in annuli near the cluster centre. This procedure is more sensitive to large-scale structures and could better characterise different sky regions. Here, we tested the sample with an inner radius of $3$ Mpc and an external one of $5$ Mpc. We do not find, statistically, significant trends in the final richness results with FAE (a median difference of $0.75$) in agreement with the initial method. 

\begin{figure}
    \centering\includegraphics[width=\linewidth]{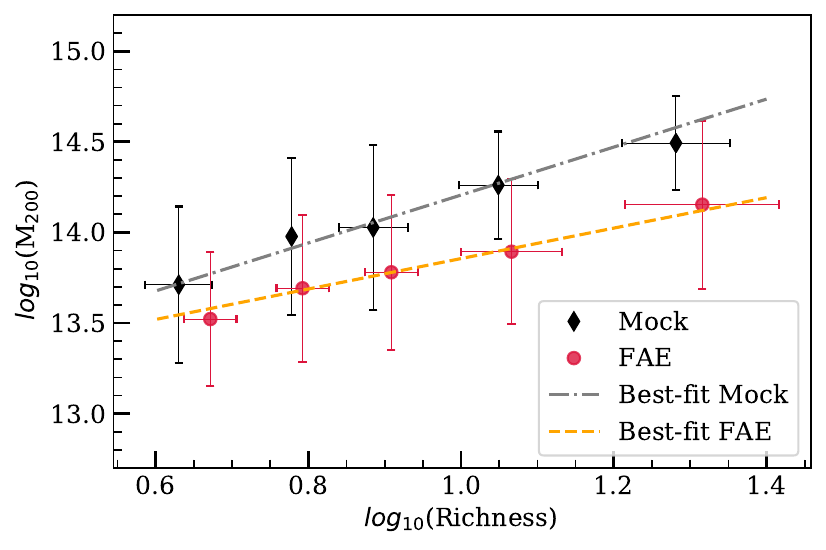}
    \caption{Richness -- mass relation obtained with the fixed aperture estimator. Black diamonds show the mean values of the mock clusters, and the red circles show the FAE results. The grey and orange lines show the linear regression for both datasets.}
    \label{fig:fix_rich_mass}
\end{figure}

We present preliminary scaling relations obtained with the above procedure in Fig.\,\ref{fig:fix_rich_mass}, and compare to the true values of the mock catalogue (more details in \S\,\ref{Data}), as orange and grey curves, respectively. Red points represent the fixed aperture richness, and black points the mock catalogue. Both datasets share the same halo masses values.
While the richness estimated by the above procedure correlates with the mass, the slope is rather flat given the scatter ($0.84\pm 0.12$) in comparison to the mock catalogue ($1.32\pm 0.11$), thus making it an uncompelling proxy. The main reason we found for this flatness is the fixed radius that does not take into consideration differences in the sizes of clusters of different masses.

\subsection{An adaptive membership estimator (AME)} \label{sec:AME}

\begin{figure}
    \centering
    \includegraphics[width=\columnwidth]{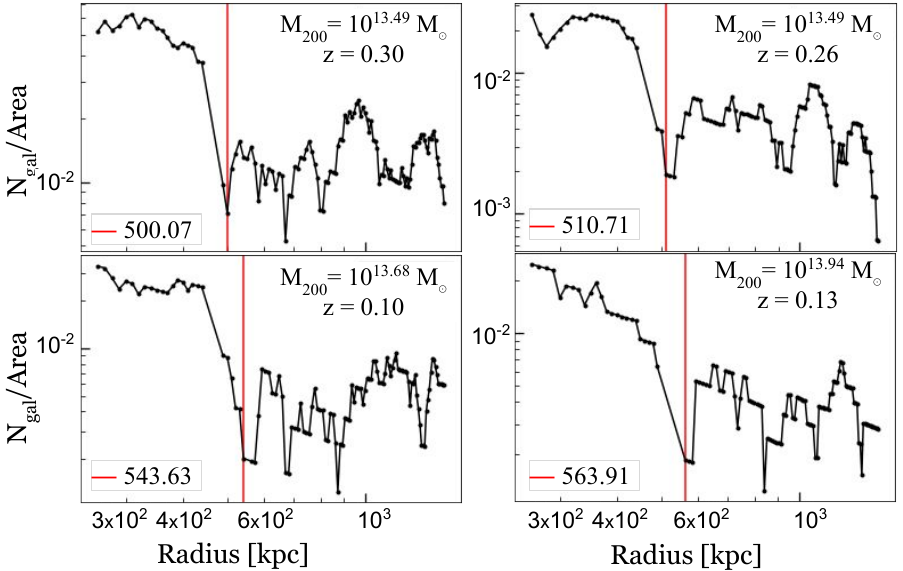}
    \caption{Example cases of the radial distribution of galaxies and $R_c$, as the red vertical line, highlighting the detected discontinuity in the central density.}
    \label{fig:rc}
\end{figure}

We propose an adaptive aperture procedure to reduce the aperture effect on the richness and keep the membership and richness estimation as data-based and assumptions-free as possible. The inputs are the galaxy catalogue projected positions, photo-z PDFs, and the cluster position in the (2+1)D space: z$_{\text{cl}}$, RA$_{\text{cl}}$ and Dec$_{\text{cl}}$. In this work, we assume the PDF as a Gaussian distribution, centred on the photometric redshift, developed to reproduce the expected values in the S-PLUS survey. In the following, we present our main algorithm steps' and details.

\begin{enumerate}
\item Remove obvious non-members by cutting out all galaxies outside a radius of $2.5$ Mpc in the plane of the sky and with $|z_{phot,i} - z_{cl}|>3\sigma_{MAD,z}(1+z_{cl})$. 
\item Calculate the galaxy density profile.  A core radius ($R_c$) will be defined as a break or ``knee'' in this profile. Estimate $\mathcal{R}$ (Eq. \ref{eq:rich_fix}) within this radius.
\item Draw a photo-z PDF-based random redshift value for each galaxy within $R_c$.
\item Calculate the sample velocity dispersion with the drawn redshifts after a $3\sigma$ clipping process. 
\item Run HDBSCAN \citep{Campello2014} in the 2D space using the remaining galaxies. Input parameters are galaxy positions and $\mathcal{R}(R<R_c)$ as the minimum cluster size parameter. The structure with the most galaxies is assumed to be a primary counterpart of the cluster in question. Minor groups are candidate substructures.
\item Repeat N times steps $3$--$5$. The probability of each galaxy being a member is the number of times that the galaxy is included as a member over N:  $P_{mem} = N_{mem} / N$. 
\end{enumerate}

In step (i), we try to ensure that no galaxies with reasonable possibilities of being gravitationally bound cluster members are excluded from the putative member pool.

Consistent with our data-oriented process, we let the radial distribution of galaxies define a projected space aperture (R$_c$) for the remaining analysis. We detect a discontinuity on the radial density profile gradient starting from the centre in step (ii). This behaviour is expected to happen when our initial sample stops being cluster dominated to be field dominated.

The projected density profile is calculated in the annulus, from the centre on, in overlapping steps of $10$ kpc and with a log-scaling width of $50$--$200$ kpc. Survey boundaries are carefully considered when estimating the areas using the Monte Carlo method. This process also provides a powerful approach to dealing with possible masked survey areas. The core radius, $R_c$, is defined as a sudden drop/break in density (e.g., a factor of 2 within steps). 

To identify this ``break'' point, we use the Kneedle algorithm \citep{Satopaa2011}. In short, the code identifies a local minimum by accounting for the difference between the density profile and a straight line connecting its initial and final points. $R_c$ is then the first (more central) detected local minima. Galaxies outside $R_c$ are discarded at this point. 

Figure \ref{fig:rc} displays some examples of the radial distribution and the ``break'' point (red vertical lines). Despite the noisy behaviour for larger radii, the central discontinuity is pronounced. We tested different binning (width of the annuli radius) and steps (increment of the smaller annuli radius). Larger steps can introduce a positive difference of $30$ to $60$ kpc over the entire range of radii. For binning, a more refined scheme does not produce a significant bias. The expectation is that the signal-to-noise contrast between field and cluster galaxy populations is maximum within this radius. In \S\,\ref{sec:rc} we discuss how $R_c$ scales with $R_{200}$.

\begin{figure}
    \centering
    \includegraphics[width=\linewidth]{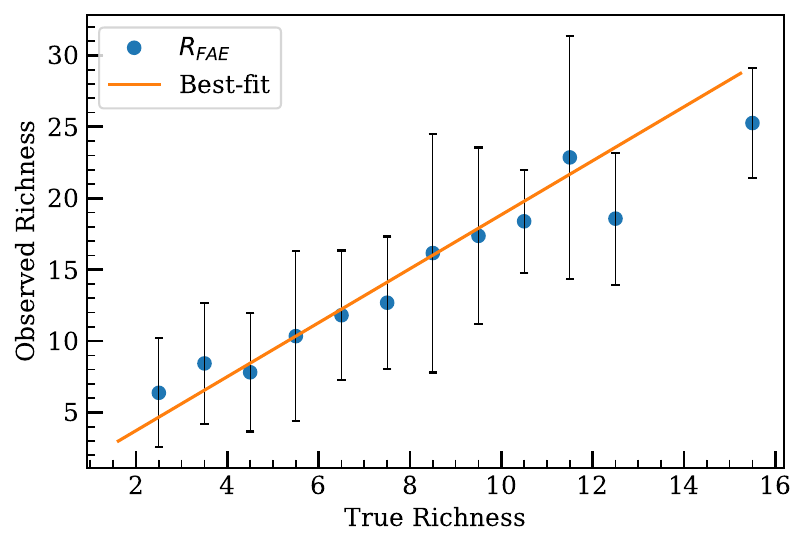}
    \caption{Relation between the number of true mock cluster members and the mean values obtained with the FAE method applied over $R_c$. The orange line represents a linear regression. This relationship can be used to infer the minimum number of neighbours to run HDBSCAN.}
    \label{fig:rich_mock}
\end{figure}

In steps (iii) and (iv), we use the photo-z PDF instead of its point estimation. We draw a value for each galaxy in the remaining sample, mimicking the realization of an ideal redshift measurement. 

After these clippings in the (2+1)D space, we still expect the sample to be somewhat contaminated. Due to the quality of the photometric redshift, tests with true member galaxies only produced a similar value in velocity dispersion. This analysis indicates that even relying on an interactive approach relating a mock quantity, e.g., richness, with the expected velocity dispersion would not produce better results. Applying a more rigorous cut ($1\sigma$) may solve this issue for data with a well-behaved PDF since the limit becomes smaller than the redshift uncertainty itself. However, in the case of a more realistic distribution, this procedure can eliminate a meaningful part of the information. Nevertheless, despite the limitations of the code, we can obtain accurate results, as discussed below.

As a final step (v), we use the density-based clustering algorithm HDBSCAN\footnote{Hierarchical Density-Based Spatial Clustering of Applications with Noise.}. It connects points in a space based on proximity, deeming more isolated ones as interlopers.   

\begin{figure*}
    \centering
    \begin{minipage}{\linewidth}
        \includegraphics[width=0.33\linewidth]{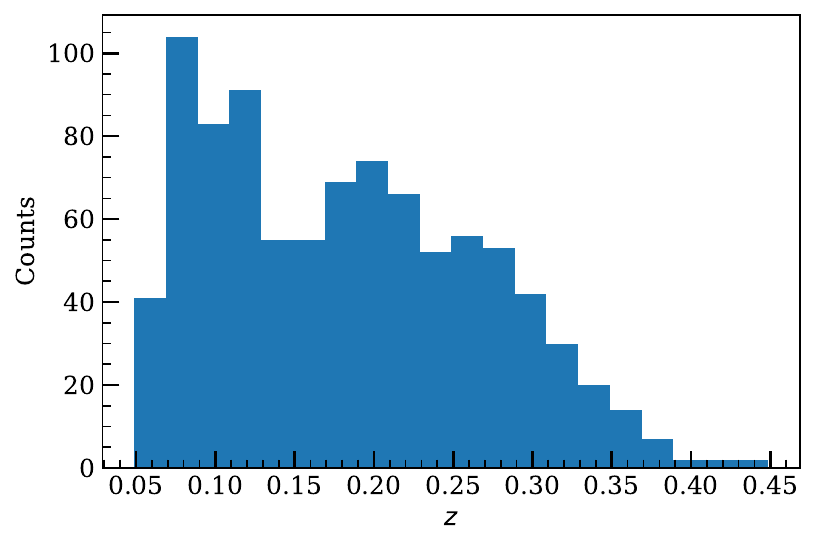}
        \includegraphics[width=0.33\linewidth]{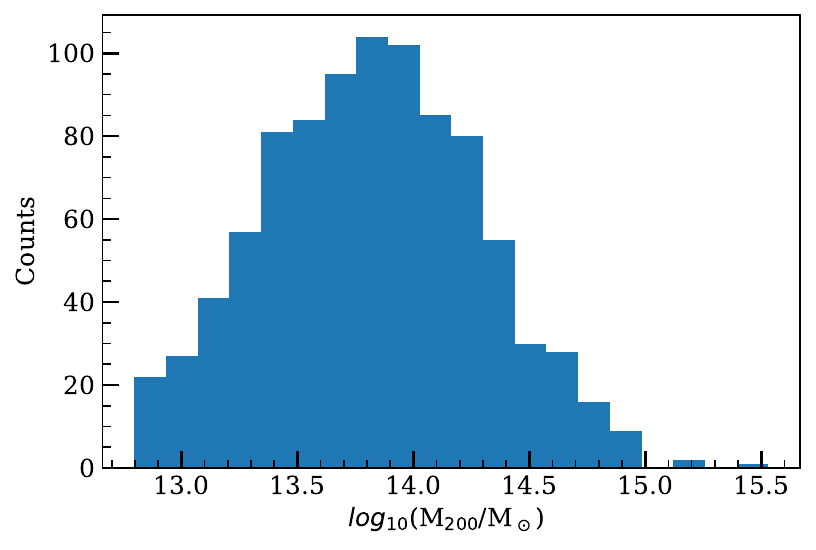}
        \includegraphics[width=0.33\linewidth]{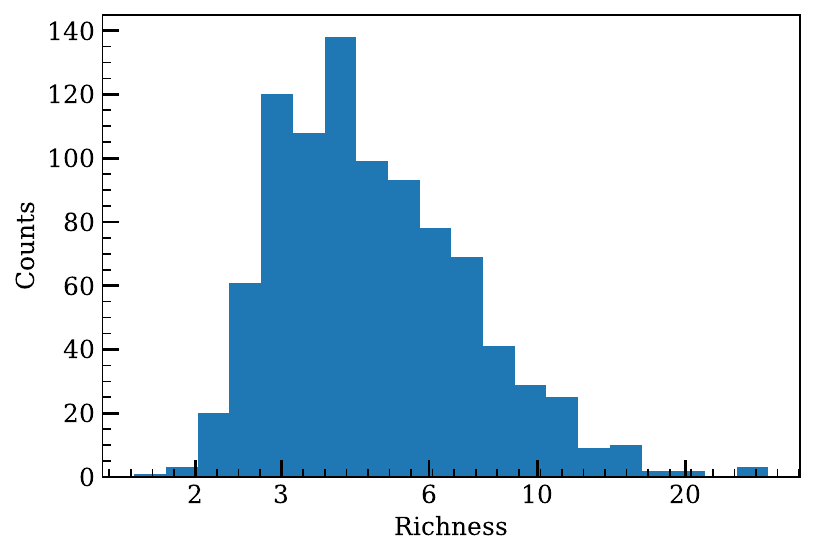}
    \end{minipage}
    \hfill
    \begin{minipage}{\linewidth}
    \centering
        \includegraphics[width=0.34\linewidth]{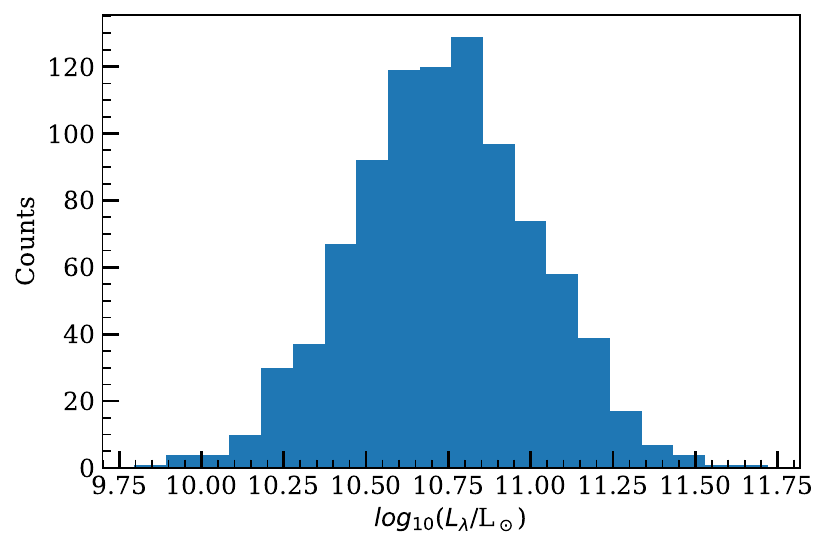}
        \includegraphics[width=0.34\linewidth]{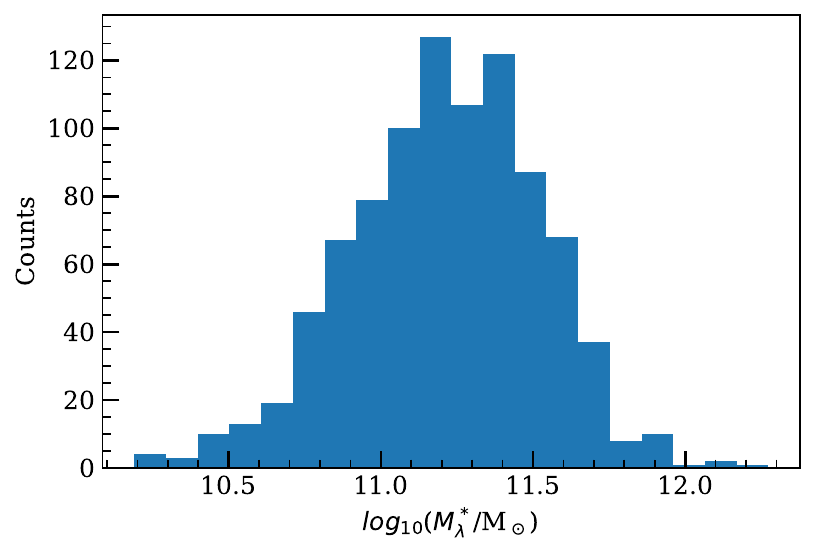}
    \end{minipage}
    \caption{Top panels: From left to right: redshift, mass and richness distribution for the mock galaxy cluster catalogue. Richness is estimated using the sum of probabilities ($P_{\rm mem}$) for each galaxy to belong to a cluster.\,Bottom panels:\,Optical luminosity, $L_\lambda$, and total stellar mass, $M_\lambda^*$.\,Both quantities\,are\,weighted\,by\,$P_{\rm mem}$.}
    \label{fig:results}
\end{figure*}

Similar algorithms, such as the better-known DBSCAN \citep{Ester1996ADA}, have been used in astronomy for similar purposes \citep{Bhattacharya2017, Olave-Rojas2018}. DBSCAN, however, requires two parameters to run: a distance-related parameter ($\epsilon$) and the minimum number of neighbours: ({\sc min\_samples}). HDBSCAN estimates $\epsilon$ by itself, varying and integrating it into a search for the most stable value. It still needs the minimum number of neighbours as an input parameter.

To provide this input, we use the relation between the number of true members (``true richness'') defined within R$_c$, instead of a fixed aperture of $500$ kpc, and $\mathcal{R}$ (Eq. \ref{eq:rich_fix}). We show in Fig.\,\ref{fig:rich_mock} the relation between the number of true mock cluster members and the average of $\mathcal{R}(R<R_c)$ values. It can be seen that the richness values we are obtaining are almost twice the true values indicating important contamination. On the other hand, as there is a linear relation between those quantities, we can use $\mathcal{R}(R<R_c)$ to estimate {\sc min\_samples}. As the richness depends on the chosen absolute magnitude limit, the slope of this relation may be slightly higher (lower) for a shallower limit (deeper).

The final membership probability is finally defined by running the procedure above $N=100$ times and estimating the membership probability of a given galaxy by the number of times it has been selected over the total number of tries. This iteration process allows for several redshift realizations of the photo-z's PDFs, thus making full use of it.  

\vspace{-0.75cm}
\section{Results} \label{Results}

We are considering galaxies with  $M<-20.25$ for the membership sample. This way, we have a volume-limited sample up to $z=0.45$.
At this redshift, this luminosity is about one magnitude fainter than the characteristic $M_*$ \citep{Puddu2021}. We calculate the absolute magnitudes simply by the relation with apparent magnitude,\,$M=m$ $-5{\rm log}_{10}(d_{\rm Mpc})-25$, where $d_{\rm Mpc}$ is the luminosity distance in Mpc.

In Fig.\,\ref{fig:results}, we show the main products of this exercise. The redshift and mass distributions of the mock clusters, plus the main mass proxies obtained through the probabilistic membership method:
\begin{align}
        &{\rm Richness}           & &\mathcal{R} = \sum_i P_{{\rm mem},i} \\
        &{\rm Optical ~luminosity} & &L_\lambda =  \sum_i L_i P_{{\rm mem},i} \\
        &{\rm Stellar~ mass}       & &M^*_\lambda = \sum_i M^*_i P_{{\rm mem},i}
\end{align}
These values depend on the galaxy's P$_{\text{mem}}$, which include only $1.1\%$ of the total galaxy population in the simulation for $M<-20.25$. The total fraction of true galaxies associated with a cluster, known from the simulations and without any cuts, is $1.75\%$. Applying thresholds for mass, redshift, and the minimum number of true associated galaxies, this value drops to $0.48\%$. 
The discrepancy between the two percentages for P$_{\text{mem}}$ and the true members highlights the contamination of our sample. 
We discuss in \S\ref{Individual Memberships} the P$_{\text{mem}}$ completeness and purity results, as well as different cluster mass and redshift ranges, to understand the source of the contamination.

\subsection{Physical meaning of \texorpdfstring{R\textsubscript{c}}{Rc}} \label{sec:rc}

Galaxy probabilities and richness estimates are typically measured within some specific aperture radius related to the cluster's physical size. Common choices are the radius within which the density is $200$ ($500$ or other) times the critical density of the Universe. 

For the results using the adaptive membership estimator (AME), assuming the radius in which HDBSCAN will act is also necessary. This step is done by searching for a break ($R_c$) in the number density profile of the cluster, as described above (Section\,\ref{Richness_code}).
A statistical comparison between the mock-defined $R_{200}$ and $R_c$ reveal that $R_c$ is $0.60$ of $R_{200}$, having a scatter of $\sigma_{MAD} = 0.16$.
Because $R_c$ does not represent a very restrictive criterion, it also can be interpreted as a physical size of the cluster, close to $R_{500}$. 

\subsection{Sanity check}

To ensure that the richness code works properly, we perform a sanity check with a catalogue of random coordinates and redshifts.
The random catalogue comprises $1000$ galaxy clusters with sky coordinates and redshifts drawn using the python function \textit{random.uniform}. The limits are defined by the boundaries of the mock catalogue. For $R_c$, we cannot apply a similar random distribution. Because groups are usually described with a smaller radius and are more numerous in counts, the function that describes the radii in the function of the structure mass follows an exponential behaviour. We therefore use $R_c$ results from the mock cluster catalogue to derive a probability distribution function. We then randomly distribute values that reproduce the data.

\begin{figure}
    \centering
    \includegraphics[width=0.49\columnwidth]{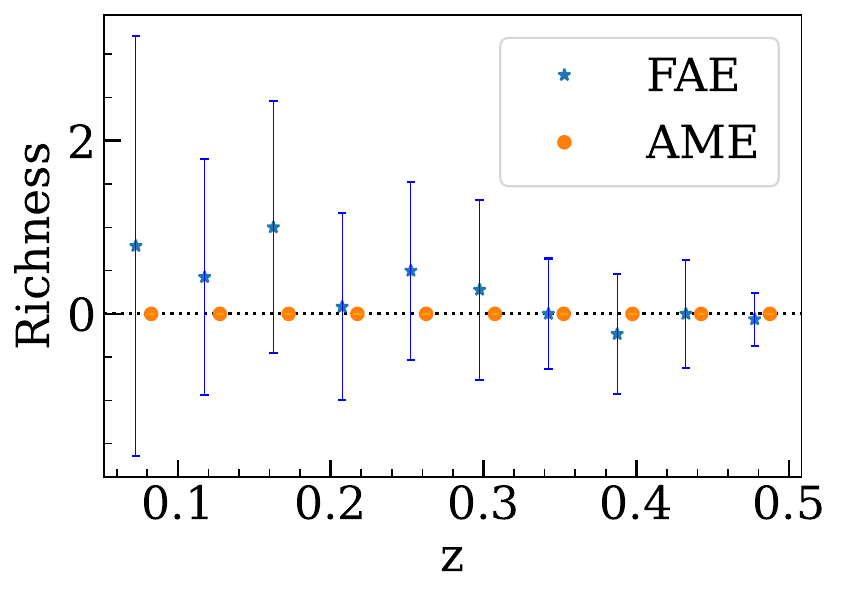}
    \includegraphics[width=0.49\columnwidth]{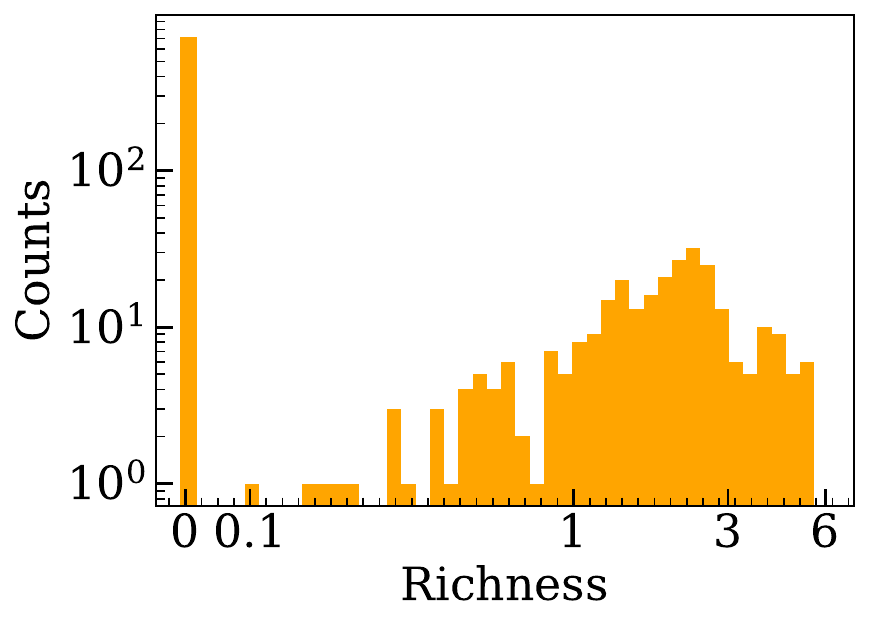}
    \caption{Left panel: Richness values for randomly distributed sky coordinates and redshifts. Values were obtained for both fixed aperture (FAE) and adaptive membership estimators (AME), represented by blue stars and orange circles, respectively. Right panel: Histogram of richness obtained with AME in logarithmic scale. Only $240$ random coordinates have a significant value (greater than $1$).}
    \label{random_richness}
\end{figure}

For each random cluster, we estimate richness values for both fixed aperture and adaptive estimators. The results are presented in Fig.\,\ref{random_richness}.
As expected, the sanity check shows median values near zero, with $\sigma_{MAD}\sim 1.05$ for the FAE method. This result indicates that richness values lower than $1.1$ may include random superpositions\footnote{For further analysis on robust decontamination process, we suggest a procedure developed by \cite{Klein2017}, on which the authors discuss an estimator that allows removing cluster candidates below a given threshold.}. Note that we have some $\mathcal{R}_{FAE}$ below zero. This behaviour occurs when the sum of the probability of the background galaxies is greater than the sum of galaxies around the fixed radius, indicating a region of galaxy density lower than the mean value.

For the AME, values are concentrated over the zero line without $\sigma_{MAD}$ error bars. However, this does not mean the entire $1000$ random points return zero. In this case, we hit a significant structure ($R_{AME}>1$) $240$ times by chance. These chance detections though are insufficient to account for the statistics (right panel of Fig.\,\ref{random_richness}).  

Another important discussion for the AME is the minimum cluster size obtained by the linear fitting. As the parameter depends on the fixed aperture estimator, values with $\mathcal{R}_{FAE}<\sigma_{MAD}$ return {\sc min\_samples} $\sim 0.6$; However HDBSCAN only works with {\sc min\_samples} $> 2$. If we simply round the values, we force the code to find at least $2$ bound galaxies. This circumstance introduces a bias of at least $3$ for the random detections at $z=0.1$, which decreases for higher redshifts. To avoid this situation, we introduce a $\mathcal{R}_{FAE}$ threshold, where $\mathcal{R}_{FAE}<\sigma_{MAD}$ HDBSCAN is unable to run, resulting in $\mathcal{R}_{AME}=0$.

\subsection{Richness significance}

The main goal of this study is to develop a methodology that produces an optical mass proxy with low scattering using photometric information. Additionally, we aim for a method that can be applied to any cluster catalogue, ranging from groups to galaxy clusters, without relying on strong modelling hypotheses.
To use optical richness as a proxy for scaling relations, we need to be able to quantify the richness with good accuracy. To check its significance, we can perform an analysis in richness bins.

\begin{figure}
    \centering
    \includegraphics[width=\columnwidth]{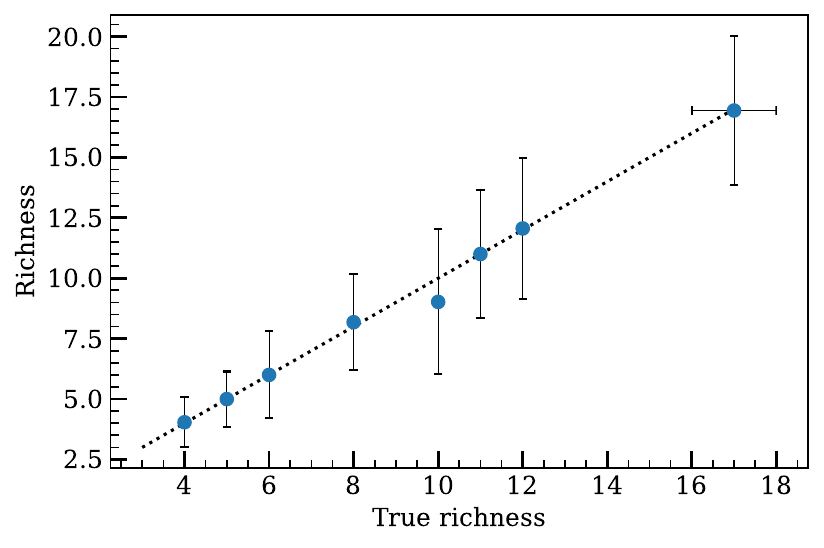}
    \caption{Richness significance in comparison between the richness calculated with the adaptive membership estimator and the ``true'' richness given by the mock. Each true richness bin contains at least $10$ clusters. The black dotted line shows the one-to-one relation.}
    \label{Rich_significance}
\end{figure}

As we know the true richness of the sample, we can compare the average values given by the mock with the richness calculated by the adaptive membership estimator ($\mathcal{R}_{AME}$). The true richness is given as the sum of the galaxies identified by the simulations located within $R_c$ from the cluster centre. 
Fig.\,\ref{Rich_significance} presents the agreement between both quantities. A black dotted line shows the one-to-one line. The residual value between both quantities, ${\rm True\, richness} - {\rm \mathcal{R}_{AME}}$, is $-0.011 \pm 0.119$. This low and unbiased result reassures the quality of our richness estimate. We require that each bin includes at least $10$ clusters.

\subsection{Mass-Observable relations}

The main difficulty in using galaxy clusters for cosmological studies is the measurement of their masses for each object. A workaround is to correlate mass with other observational properties, such as optical richness, X-ray luminosity, or total stellar mass. This relationship between mass and observable is usually calibrated using a limited sample of objects and then extended to the full catalogue. For example, lensing surveys consider the full sample in stacked analyses. An interesting mass proxy should present a low intrinsic scatter to produce robust mass estimates.

Our AME returns for each galaxy the probability of being physically bound to a certain cluster. Using this capability, we can calculate other proprieties, weighted by the membership, to characterize the cluster sample. 
For example, with the magnitude information in the $r$-band we can estimate the total optical luminosity, $L_\lambda = \sum L_i\,P_i = \sum 10^{0.4[4.42 - M_i]}\,P_i$. The solar absolute magnitude is represented by $4.65$ in r-band \citep[$SDSS_r$, ][]{Willmer2018}, and the $i$-th galaxy absolute magnitude in the same band by $M_i$. 
The mock galaxy's stellar mass can be used to derive the total stellar mass, following the same procedure as described above, $M^*_\lambda = \sum M^*_i\,P_i$, with $M^*_i$ as the $i$-th galaxy stellar mass. 

A similar approach can be applied to derive the mock quantities. In this case, $P_i$ is always one for the true galaxy members and zero otherwise.

\begin{figure}
    \centering
    \begin{minipage}{\columnwidth}
      \centering
      \includegraphics[width=\columnwidth]{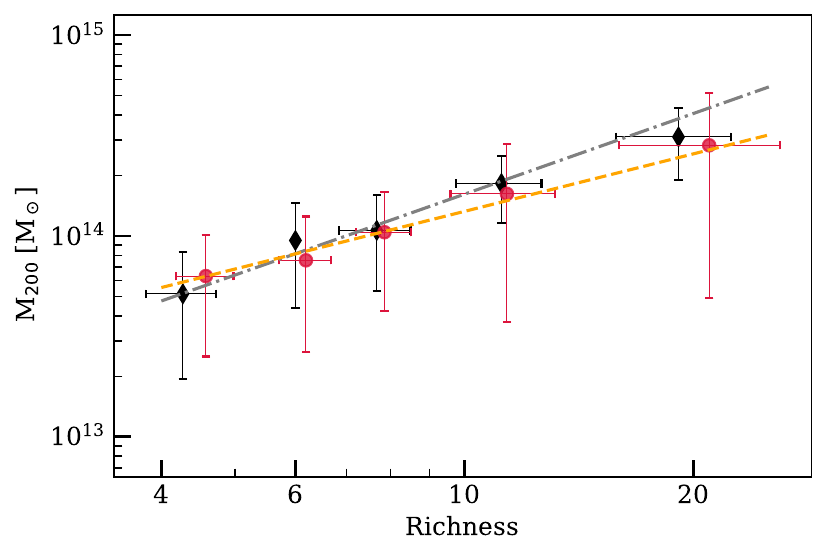}
    \end{minipage}
    \hfill
    \begin{minipage}{\columnwidth}
      \centering
      \includegraphics[width=\columnwidth]{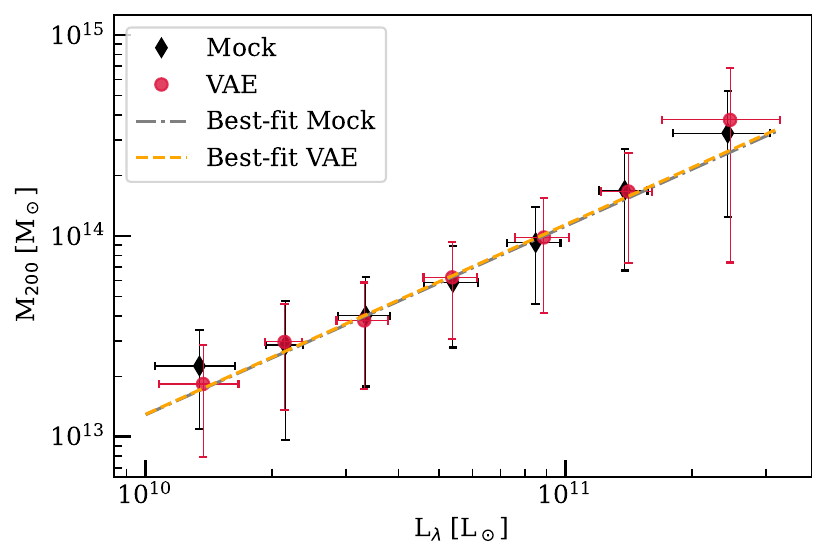}
    \end{minipage}
    \hfill
    \begin{minipage}{\columnwidth}
      \centering
      \includegraphics[width=\columnwidth]{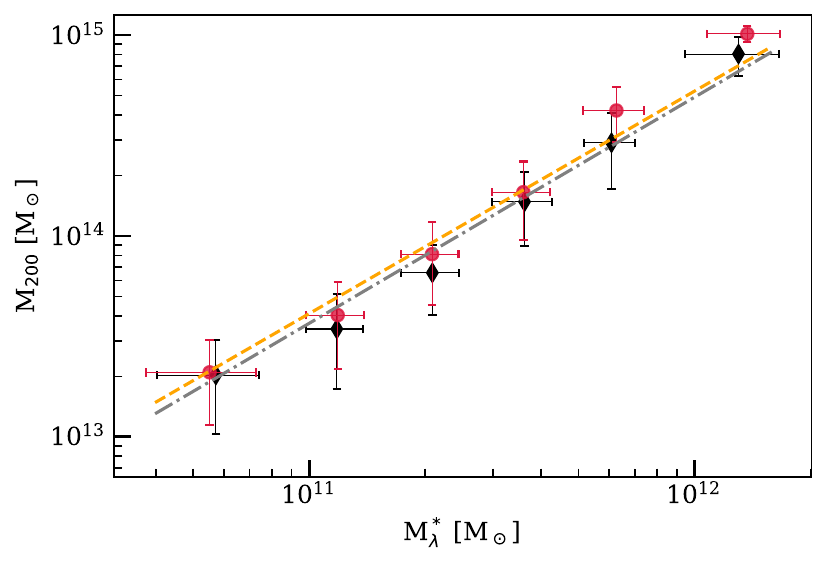}
    \end{minipage}

    \caption{Scaling relations between mass and optical proxies. We highlight the median values in proxy bins for both the adaptive membership estimator (as red dots) and mock values (as black diamonds) and the linear regressions, orange and grey lines, respectively. Coefficient values can be found in Table\,\ref{tab:fit_values}. Top panel: Mass-Richness relation. Middle panel: Mass-Optical Luminosity. Bottom panel: Mass-Stellar mass.}
    \label{fig:M200_qtd}
\end{figure}

With our mass proxies, we can derive the scaling relations. The linear regression is done by {\sc linmix} \citep{Kelly2007}. This algorithm uses a Bayesian approach and accounts for errors in both parameters for the minimization process, masses and proxies, which is ideal for real data. The relation is modelled as, 
\begin{equation} \label{eq:reg}
    \log_{10}(M_{200}) = \alpha + \beta\,\log_{10}\left(\frac{O}{O_{piv}} \right) \pm \epsilon
\end{equation}
where coefficients are represented by $\alpha$ and $\beta$, $O$ is the mass proxy, $O_{piv}$ is a pivot value corresponding to the mean value of the mock proxy, and $\epsilon$ is the intrinsic random scatter about the regression. Table\,\ref{tab:fit_values} shows the resulting best-fitting parameters for both mock (marked with a symbol $\dagger$) and estimated values from the AME. The relation between median values of mass and Richness (top), $L_\lambda$ (middle), and $M^*_\lambda$ (bottom panels), separated in proxy bins, can be found in Fig.\,\ref{fig:M200_qtd}. Black diamonds represent mock and red circles represent the estimator results.

\begin{table} 
    \caption{Best fitting values of the linear regression for richness ($\mathcal{R}$), optical luminosity ($L_{\lambda}$) and total stellar mass ($M^*_{\lambda}$). The model mass-observable is described by Equation\,\ref{eq:reg}. $O_{piv}$ is always related to the mock proxy.} $L_{\lambda}$ and $M^*_{\lambda}$ are given in units of L$_\odot$ and M$_\odot$. Mock results are identified by $\dagger$ after the observable proxy.
    \centering
    \begin{tabular}{c c c c c}
    \hline
    Proxy & $\alpha$ & $\beta$ & $\epsilon$ & $O_{piv}$\\
    \hline
    $\mathcal{R}\dagger$ & $13.85 \pm 0.02$ & $1.32 \pm 0.07$ & $0.180 \pm 0.009$ & $5.4$\\
    $\mathcal{R}       $ & $13.86 \pm 0.02$ & $0.95 \pm 0.10$ & $0.181 \pm 0.009$ & ''\\
    $L_{\lambda} \dagger$ & $13.89 \pm 0.01$ & $0.94 \pm 0.05$ & $0.141 \pm 0.007$ & $6.7\times10^{10}$\\
    $L_{\lambda}$         & $13.89 \pm 0.01$ & $0.95 \pm 0.05$ & $0.151 \pm 0.007$ & ''\\
    $M^*_{\lambda}\dagger$ & $13.92 \pm 0.01$ & $1.12\pm 0.04$ & $0.092 \pm 0.005$ & $2.4\times10^{11}$\\
    $M^*_{\lambda}$        & $13.99 \pm 0.01$ & $1.12 \pm 0.03$ & $0.097 \pm 0.005$ &''\\ 
    \hline
    \end{tabular}
    \label{tab:fit_values}
\end{table}

Statistically, the most scattered distribution is the richness-mass relation. The mock results for the low-mass end highlight the high intrinsic scatter for the small richness groups. The same structure with $\log_{10}(\mathcal{R}) = 0.6$ ($\mathcal{R} = 4$) can have a halo mass between $10^{13}$ and $10^{13.9}$ M$_\odot$. Besides the different low richness end, both (mock and adaptive estimators) linear regressions present the same behaviour. The observed intrinsic scatter is $\sigma_{\log_{10}(M|\mathcal{R})} = 0.181\pm0.009$ dex, a value similar to the best-case scenario, i.e. simulations.
The total optical luminosity is a valuable parameter with median $\sigma_{log_{10}(M|L_\lambda)} = 0.151$ dex, compared to $\sigma_{log_{10}(M|L_\lambda)} = 0.141$ dex from simulations. This residual scatter observed between mock and AME results is consistent with the intrinsic scatter of the relation $\mathcal{R}\dagger$-$\mathcal{R}_{AME}$, of $0.014$. 
Considering the mass range amplitude, similar behaviour is observed as $L_{\lambda}$ depends mainly on magnitudes and galaxies probabilities. We see a small deviation for lower luminous structures, probably due to external contamination, that slightly increases the intrinsic scatter in relation to the best-case scenario.  
$M^*_{\lambda}$ is an interesting option with the lowest intrinsic scatter. It provides a galaxy cluster candidate's characterization regarding physical properties, such as stellar mass. These results may be overly-optimistic since we use the exact values of the stellar masses. In optical surveys, scatter may be introduced by the inference method. The same behaviour found in $L_{\lambda}$ for the low end is seen here. We also observe a small difference in $\beta$ that introduces a $\sim 0.01$ gap.

\subsection{Perturbing the cluster centre and redshift} \label{real-life}

We can test the code robustness by applying small variations in redshift and coordinates in the mock cluster catalogue that mimics expected disturbances found in galaxy cluster detections using observational data. 

\cite{Werner2022} discuss the application of the density-based algorithm PZWav \citep{Gonzalez14} for detecting clusters from S-PLUS DR1 and analyze the algorithm performance by comparing the outputs with the same mock catalogue used in our work. Due to the PZWav computational approach that detects substructures by the spatial distribution of the galaxies and estimates the redshift also based on the surrounding galaxies, the variations found in centre distances and redshifts are notably small. The mean radial difference is $10$ kpc with a standard deviation of $12$ kpc, and the mean redshift difference is  $0.6\times 10^{-3}$ with $\sigma = 8.8\times10^{-3}$.  

We now test different scenarios that include the mean value added to the $1 \sigma$ ($\Delta R = 22$ kpc, $\Delta z = 0.009$), $2 \sigma$ ($\Delta R = 34$ kpc, $\Delta z = 0.018$) and $3 \sigma$ intervals ($\Delta R = 46$ kpc, $\Delta z = 0.027$) as the peak of a 2D Gaussian distribution \citep[such as][for miscentering]{Johnston2007,Vitorelli2018}, and compare the obtained richness results with the centralized ones.
For the $\Delta R$ variations, we do not use a preferential frame for the displacement. We draw instead a random value between $0$ to $360$ degrees that orient the offset. 

While the radial variation is always positive, $\Delta z$ can be applied as a positive or negative offset. For this, we use normal distributions, centred on $\Delta z$ with proportional $\sigma_{Norm}=1$, $2$ or $3\sigma$ that randomly that adds or subtracts from the cluster redshift. 

In Table\,\ref{tab:real-life}, we show the median relative errors, and $\sigma_{MAD}$, between the richness results for $\Delta R$ and $\Delta z$ separately. As there were no significant trends through the redshift range, we present only one result containing all the galaxy clusters for each scenario.

For the centre distance, there is no effective deviation in richness results for small $R$ offsets, i.e., the median relative errors are consistent with zero. This analysis shows the code is robust to centre variations lower than $R<46$ kpc. Investigating further, we found that the median relative error is below $-2\%$ until $\Delta R> 250$ kpc, increasing to $-35\%$ for $\Delta R> 500$ kpc and $-50\%$ for $\Delta R> 750$ kpc. Note that the negative sign represents an underestimation of the richness. Those values indicate that the code is sensitive to differences in the galaxy density at cluster outskirts compared to the central distribution and the choice of the background estimate using the farthest galaxies as contamination points.

For the redshift offsetting, we observe a decrease (however, within error bars) in the richness measurements, going from $-8\%$ for $2\sigma$ to an $-18\%$ deviation for $3\sigma$. Such difference as large as the photo-z uncertainty, causes it to exclude a significant contribution from galaxies near the centre of the cluster. This analysis shows that the code is robust to both centre and redshift variations.

\begin{table}
    \centering
    \caption{The median relative error between richness results before and after the $\Delta R$ and $\Delta z$ offsets, considering a typical mean value added to the $1 \sigma$ ($\Delta R = 22$ kpc, $\Delta z = 0.009$), $2 \sigma$ ($\Delta R = 34$ kpc, $\Delta z = 0.018$) and $3 \sigma$ interval ($\Delta R = 46$ kpc, $\Delta z = 0.027$).}
    \begin{tabular}{c c c}
        \hline
                   & $\Delta R$ [kpc]         & $\Delta z$ \\
       \hline
       $\bar{O}+1 \sigma$  &  $0 \pm 0.010$ & $-0.02 \pm 0.05$ \\
       $\bar{O}+2 \sigma$  &  $0 \pm 0.011$ & $-0.08 \pm 0.10$ \\
       $\bar{O}+3 \sigma$  &  $0 \pm 0.013$ & $-0.18 \pm 0.18$ \\
       \hline
    \end{tabular}
    \label{tab:real-life}
\end{table}

\begin{table}
    \centering
    \caption{Comparison between AME richness results with more realistic photo-z PDFs as Student’s t-distribution and different fractions of bi-modality. }
    \begin{tabular}{c c c}
        \hline
        Photo-z PDF & fraction  & median relative errors \\
       \hline
       t-distribution & $100\%$ & $-0.02 \pm 0.03$ \\
       \hline
       \multirow{4}{*}{bimodality} & $10\%$  &  $0.01 \pm 0.02$ \\
        & $25\%$  &  $-0.02 \pm 0.05$ \\
        & $50\%$  &  $-0.15 \pm 0.06$  \\
        & $100\%$  &  $-0.31 \pm 0.08$ \\
       \hline
    \end{tabular}
    \label{tab:gaussian}
\end{table}

\subsection{More realistic photo-z PDFs}

In this work we approximate the photo-z PDFs as well-behaved Gaussian distributions. However, PDFs from observations might be more complex, often presenting long tails or secondary peaks.
As the photo-z PDFs play a central role in defining the richness, we also test the effects of the non-Gaussianity in our sample.

To address the effects of the long tails, we describe the galaxy sample with Student's t-distributions. We centred it on the galaxy photo-z central value, with width $\sigma_{MAD,z}$ (\S\,\ref{Data}). We adopt the degrees of freedom parameter $\nu =1$ to make the wings much broader than the Normal ones. This test produced similar results in terms of richness, with a relative error of $2\%$, as presented in Table\,\ref{tab:gaussian}.

To test the effect of a bimodal PDF, we took some examples from the S-PLUS database. We define that a PDF has a significant secondary peak when the distance between peaks is greater than $0.05$, and the secondary peak amplitude is at least $20\%$ of that of the primary peak. For each galaxy in the mock, we attribute an S-PLUS-like double-peaked PDF in the function of the galaxies with the closest redshift and magnitude. We tested different fractions of double-peaked PDF, such as $10\%$, $25\%$, $50\%$ and $100\%$, and found that the richness measurements are only affected for values $> 50\%$, leading to underestimates of $15\%$.

As this work is inspired by imaging surveys that use several narrow-band filters, we do not expect double-peak to be a more significant issue than our test. Surveys such as S-PLUS and J-PAS utilize a combination of filter systems that allow a better constraint of the photo-z PDFs. S-PLUS, for example, present a secondary peak fraction lower than $10\%$.

\begin{figure}
    \centering
    \includegraphics[width=\columnwidth]{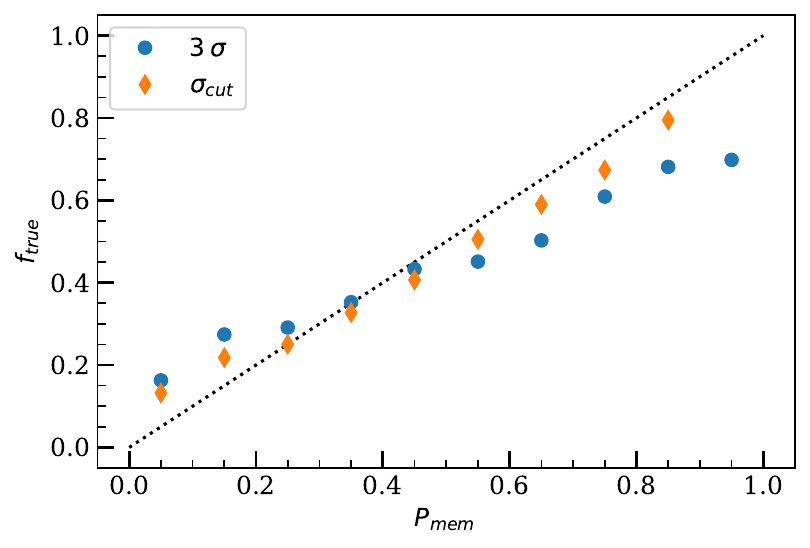}
    \caption{Galaxy membership significance as a comparison between the fraction of true members $f_{true}$ and estimated membership probabilities $P_{mem}$, for two different approaches: the default $3\,\sigma$, and a more rigorous cut $\sigma_{cut}$ based on known information from the mock.}
    \label{fig:member_significance}
\end{figure}

\section{Individual Memberships} \label{Individual Memberships}

In this section we present a view of the membership significance, compare the galaxy probabilities with the fraction of true members, discuss completeness and purity in terms of galaxy probabilities in different redshift and cluster mass ranges, and compare the cluster position and redshifts with the ones estimated by the selected member galaxies alongside the ones obtained using a cluster finder.

\subsection{Significance}

In the previous section we based our proxy estimator on the individual membership probabilities, $P_{mem}$. Here we will investigate the real meaning of this number using an approach similar to \cite{Castignani2016} and \cite{Lopes2020}. The idea is to compare the fraction of true members ($f_{true}$) given by the mock and the estimated $P_{mem}$. The galaxy data is binned in $P_{mem}$, and $f_{true}$ is calculated for each bin. 

Here we also introduce a test case based on the already known information from the mock to observe how a more strict limit in the velocity dispersion (step iv \S\ref{sec:AME}) can improve our results. First, we run AME using only the true member galaxies over the entire cluster sample and estimate the ``real'' velocity dispersion given the photo-zs. As no significant evolution with richness is expected, we take the median value of the sample. This step produces a smaller $\sigma$ than when considering all galaxies (members and non-members) within $R_c$. We also modify the clipping process: only the galaxies within $1.5\,\sigma$ are accepted instead of default $3\,\sigma$. Finally, running AME with the new limits should produce purer results.

The resulting fractions are presented in Fig.\,\ref{fig:member_significance}. A black dotted line shows the one-to-one line, blue points represent our default model, and the orange diamonds are the $P_{mem}$ values obtained with rigorous mock-based $\sigma_{cut} = 1.5\,\sigma$ limit.  

In both cases, we detect a slightly higher fraction of true members for low $P_{mem}$. On the other hand, on the high $P_{mem}$ end the contamination is higher than expected. The combination of both effects produces a flattened curve in Fig. \ref{fig:member_significance}. 
A fraction of line-of-sight contamination is expected because photometric redshift uncertainties are one order of magnitude larger than typical cluster velocity dispersions. The average fraction is $\sim 11\%$, increasing contamination for $P_{mem}>75\%$. This behaviour indicates that some non-members with the maximum peak of the photo-z PDFs near the cluster redshifts cannot be removed with the $3\,\sigma$ clipping. Still, a more rigorous cut could reduce the contamination contribution. For example, for $\sigma_{cut}=1.5$ the average fraction of contamination is decreased to only $\sim 6\%$. However, the $\sigma_{cut}$ limit is lower than the redshift uncertainty itself, restraining $P_{mem}$ at $85\%$ and introducing a small underestimate in our optical proxies.

We highlight that contamination by interlopers still occurs even in a controlled situation where we know a priori the sample velocity dispersion. Despite the higher fraction of interlopers found using the data-driven shallower threshold, we still obtain accurate results that allow us to produce scaling relations with competitive intrinsic scatter. A better characterization of how contamination affects the purity of our sample is discussed below (subsection \ref{cp}).

\subsection{Completeness and purity} \label{cp}

Another way to quantify the robustness of our method is by analyzing the completeness (C) and purity (P). Usually, completeness is described as the fraction of true members correctly classified as a member, and purity is the fraction of true members concerning all objects selected as members.

Using the same notation of other classification studies \citep{Castignani2016, Lopes2020}, we estimate C and P as,
\begin{equation}
    C = \frac{N_{\rm true} - N_{\rm missed}}{N_{\rm true}}
\end{equation}
\begin{equation}
    P = \frac{N_{\rm selected} - N_{\rm interlopers}}{N_{\rm selected}}
\end{equation}  
where $N_{\rm selected}$ is the number of galaxies that were classified as members, $N_{\rm true}$ represents the number of true members given by the simulation, $N_{\rm interlopers}$ gives the number of false positive objects (wrongly classified as a true member), and $N_{\rm missed}$ are the false negative objects (a true member with negative output).

\begin{figure}
    \centering
    \includegraphics[width=\columnwidth]{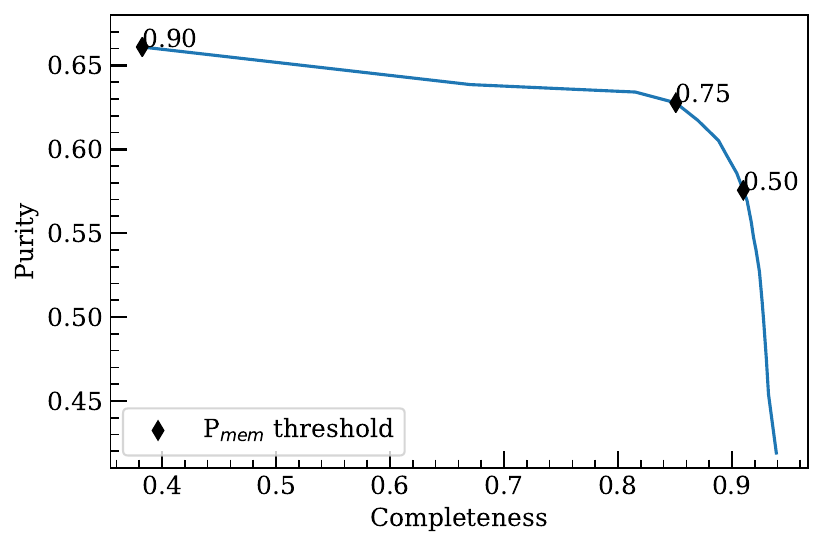}
    \caption{Completeness and purity values applying different $P_{mem}$ limiting values. We highlight $P_{mem}>0.50$, $0.75$, $0.90$.}
    \label{fig:comp_purity}
\end{figure}

We can explore variations in both purity and completeness by selecting galaxies that have membership probabilities higher than a given limiting value.
In Fig.\,\ref{fig:comp_purity}, we show completeness and purity as a function of different $P_{mem}$ thresholds, highlighting $P_{mem}>0.50$, $0.75$, $0.90$ in black diamonds. The corresponding results are $P = 0.58$, $0.63$, $0.66$ and $C=0.91$, $0.85$, $0.38$.
Note that increasing the limiting value higher than $0.75$ does not improve significantly (less than $3\%$) the purity. This limited improvement can be seen in the flattening of the curve in Fig.\,\ref{fig:comp_purity}. 

\begin{figure}
    \centering
    \begin{minipage}{\columnwidth}
      \centering
      \includegraphics[width=\columnwidth]{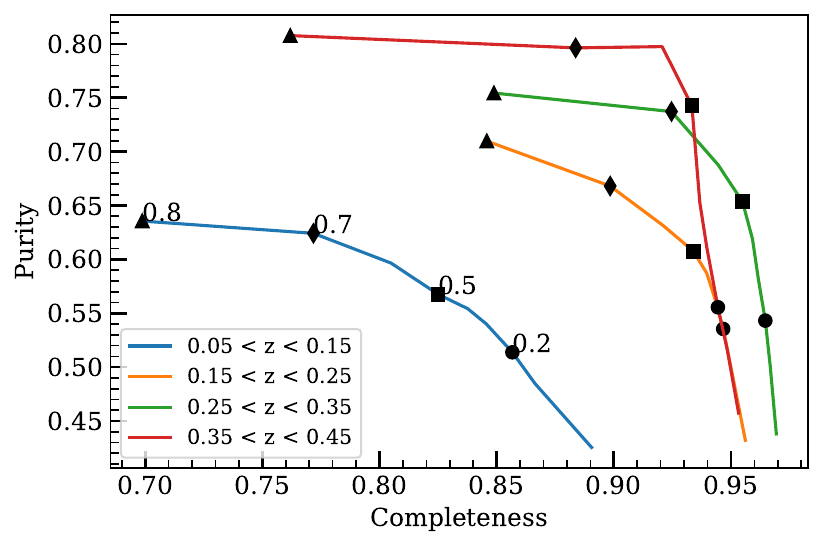}
    \end{minipage}
    \hfill
    \begin{minipage}{\columnwidth}
      \centering
      \includegraphics[width=\columnwidth]{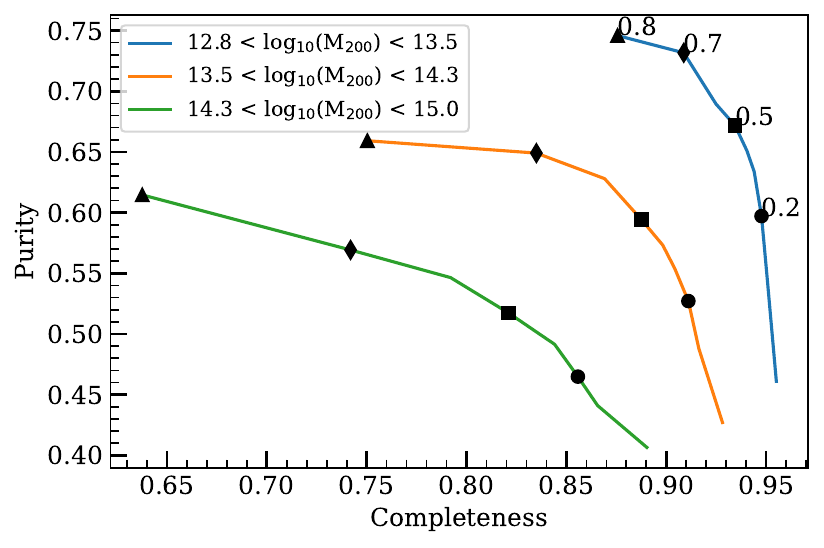}
    \end{minipage}

    \caption{Same as Fig.\,\ref{fig:comp_purity} for distinguished galaxy cluster intervals. Upper panel: different redshift ranges with steps of $dz=0.1$. Bottom panel: mass ranges with steps of $d {\rm log}_{10}(M) = 0.75$. For both, we highlight $P_{mem}>0.2$, $0.5$, $0.7$, and $0.8$ cuts as a circle, square, diamond, and triangle, respectively.}
    \label{fig:comp_purity_zm}
\end{figure}

To identify possible dependencies on cluster redshift and halo mass, we divided the cluster sample into redshift bins of $dz=0.1$ and $d {\rm log}_{10}(M) = 0.75$. Fig.\,\ref{fig:comp_purity_zm} shows the results highlighting $P_{mem}$ threshold of $0.2$, $0.5$, $0.7$ and $0.8$ as a circle, square, diamond, and triangle, respectively.
In a similar approach, \cite{Castignani2016}, using a $P_{mem}>0.2$ cut, observed mean values with only a few per cent variations ($\sim 3\%$). Taking for comparison $P_{mem}>0.2$, we see in the upper panel of Fig.\,\ref{fig:comp_purity_zm} a large difference between the $0.05<z<0.15$ (blue line) and the rest of the redshift range. This difference is $9\%$ in completeness and without a significant gain in purity ($2\%$). 
Similar behaviour is observed in completeness for the other $P_{mem}$ cuts ($\sim 12\%$), but with also an improvement in purity ($\sim 5$). 
Smaller variations are observed within $0.15<z<0.35$, with increasing improvements in both $P$ and $C$ ($\sim 5$, $\sim 2$). The red line for $0.35<z<0.45$ shows a decrease in completeness for $P_{mem}>0.8$ of $8.7\%$, with the highest purity values ($81\%$).  

For mass evolution, we observe a decrease in both C and P for more massive objects. For $P_{mem}> 0.2$, C shows a change of $-4.5\%$ and P a $-6.6\%$.
Changes are larger for high $P_{mem}$ cuts ($C = -5.7, -8.3, -11.9$, and $P = -7.7, -8.1, -6.5$ for $P_{mem}> 0.5, 0.7, 0.8$, respectively). 
We conclude that this behaviour does not originate from poor clusters, but is instead due to contamination in the cluster's outskirts from field galaxies. Besides the code's ability to remove spatially dispersed points, the ones too close to the real galaxy members are still considered. 
Massive clusters are usually populated with a larger sample of true galaxies over less concentrated areas than the group mass scale \citep{Comerford2007, Merten2015}. This effect means that the code will be more susceptible to accepting contributions from other field galaxies nearby, increasing the contamination. This lower density within the cluster core may lead to neglect of the outskirts. Less massive objects tend to be more concentrated, leading to more accurate results.

Studies with probabilistic membership assignments typically select a probability threshold to account only for reliable cluster members. In this work, we consider the contribution of all galaxies for the richness estimation. A more careful approach should be considered in the case of single cluster analyses due to the non-negligible fraction of outliers. However, for large samples, we present statistically robust results with low median differences and purity and completeness that are comparable to other estimators \citep[$P_{mem}>0.5$ $C=92\%$ $P=69\%$, $P_{mem}>0.5$ $C = 68\%$ $P=69\%$,][respectively]{George2011, Castignani2016}.

\subsection{Clusters position and redshift from \texorpdfstring{P\textsubscript{mem}}{Pmem} } \label{new_centre}

\begin{table}
    \centering
    \caption{Centre and redshift offset values between known mock/PZWav clusters and the ones obtained by galaxies accepted as a member ($P_{mem}>0$) around its centre.}
    \begin{tabular}{c c c}
        \hline
                   & $\Delta R$ [kpc]         & $\Delta z$ \\
       \hline
       O$_{mock}$-O$_{gals}$ &  $63.98 \pm  34.76$ &  $-0.0006 \pm 0.0049$ \\
       O$_{PZWav}$-O$_{gals} $  &  $128.38 \pm 53.61$ & $0.0013 \pm 0.0061$ \\
       \hline
    \end{tabular}
    \label{tab:new_centre}
\end{table}

Another test we can perform is to measure the difference between the redshift derived from the member galaxies ($P_{mem} > 0$) and the mock cluster, as well as the centre discrepancies.
We can take advantage of the galaxy cluster catalogue produced by PZWav for the mock \citep[described in][]{Werner2022} and compare how these values change in a real case of detection. For this analysis, we consider as a ``counterpart'' the PZWav detections that have a redshift difference with the mock cluster of $0.03$ and centre distance of $500$ kpc.  

For redshifts, we can compute the average value weighting each galaxy photo-z contribution by its P$_{mem}$, as $z_{gals} = \sum z_i \, P_{mem, i}/\sum P_{mem, i}$. This procedure ensures that galaxies weakly related to the cluster have a lower contribution in the calculations. 
We follow the procedure proposed by \cite{Castignani2016} for the centre estimates, where the authors estimate the mock clusters' barycentre by averaging the member galaxies' Cartesian coordinates.
Here, we also introduce the probability weight for galaxies with $P_{mem} > 0$. We do the same for the right ascension, $\alpha_{gals} = \sum \alpha_i \, P_{mem, i}/\sum P_{mem, i}$, and for declination. 

In Table\,\ref{tab:new_centre} we show as O$_{mock}$-O$_{gals}$ the difference between the quantity O ($\Delta R$ or $\Delta z$) for the mock catalogue and the averaged value obtained by the galaxies accepted as a member ($P_{mem}>0$) around its centre. Similarly, O$_{PZWav}$-O$_{gals}$ is the difference between the quantity O for the PZWav catalogue and the selected galaxies around PZWav centres.
Here we also average all the cluster samples due to insignificant trends through the redshift range.

For both scenarios, the redshift estimates are highly consistent within the detection and selected member galaxies, giving a total difference consistent with zero, with $\sigma_{MAD} \leq 10^{-3}$.
This result highlights the applied computational approach to calculate the redshift estimates. Both PZWav and simulations calculate $z$ by averaging the nearby galaxies.

In contrast, the centres present substantial discrepancies, which arise from selection effects. The centre estimations in the mock clusters are based on the true members assigned independently of the galaxy's centre distance. 
In our method, however, galaxies are selected within a cutoff radius $R_c$, which leads to the removing some distant true members from the analysis. We also highlight the presence of contaminating galaxies that can offset the centre depending on their $P_{mem}$. The median error is $64 \pm 35$ kpc. 
For PZWav, centres are estimated through overdensities, calculated based on the galaxy distribution weighted by the integrated PDF (similarly to eq.\,\ref{eq:cdf}, more details in \cite{Werner2022}). These overdensities may have physical sizes within $400$ to $1500$ kpc. Then, the centre estimate considers the contribution of all galaxies inside the range. 
This circumstance returns a median error of $128 \pm 54$.

We highlight that these values can be redshift dependent when considering angular sizes, on which for physical units we might under-estimate the offsets of low redshift clusters and overestimate high-z clusters. In this case, we found a variation of $1.3$ arcmin at $z<0.1$ that rapidly drops for $0.6$ arcmin at $z<0.2$ and stabilizes at $0.45$ arcmin within $0.3<z<0.45$.

\section{Summary and conclusions} \label{Conclusions}

In this work, we present a probabilistic membership assignment estimator that uses photometric parameters to derive reliable richness estimates, optical luminosity, and total stellar mass for 919 simulated structures with masses ranging from groups to galaxy clusters.  
The approach assumes minimal information about the definition of the cluster and uses the sky position within a characteristic radius ($R_c$) to select potential galaxy members.

Below, we highlight the main findings of this work:
\begin{itemize}
    \item The characteristic radius $R_c$ statistically scales as $R_c = 0.6 \,R_{200}$, with a median absolute deviation of $0.16$.
    
    \item Tests with random points distributed along the redshift of the simulations show that the FAE method returns $\sigma_{MAD}$ of $\sim 1.05$. By applying this value as a richness threshold for the adaptive membership estimator richness, $\mathcal{R}_{AME}$, correctly yielded $\mathcal{R}_{AME}=0$.
    
    \item Considering the group/cluster sample, comparisons between $\mathcal{R}_{mock}$ and $\mathcal{R}_{AME}$ produce a linear relation in median values, with a deviation of $-0.01 \pm 0.12$.
    
    \item With the probabilistic results, we successfully derive optical mass proxies that are simple, with low observational cost, and present small intrinsic scatter. 
    \subitem From the scaling relations:
    \subitem$\sigma_{log_{10}(M|\mathcal{R})} = 0.181 \pm 0.009$ dex
    \subitem$\sigma_{log_{10}(M|L_\lambda)} = 0.151 \pm 0.007$ dex
    \subitem$\sigma_{log_{10}(M|M_\lambda^*)} = 0.097 \pm 0.005$ dex. 
     
    \item We show that our adaptive estimator is robust for the small centre and redshift offsets. Displacements presented by \cite{Werner2022} produce richness variations lower than $1\%$. 

    \item Testing the use of complex photo-z probability density functions (PDF) reveals that the presence of long tails does not significantly impact the results. However, in cases where the PDF exhibits a bimodal distribution, there is a potential underestimation of richness by approximately $15\%$ when more than half the sample has bimodal PDFs. Considering that the fraction of bi-modality in the S-PLUS data is only $10\%$, we anticipate that the double-peak phenomenon is not a significant concern for our data.
\end{itemize}

Regarding the individual membership probabilities:
\begin{itemize}
    \item The fraction of true galaxy members in each $P_{mem}$ bin emphasizes the contamination for higher probability values. 
    \subitem The best agreement between purity and completeness is $P=63\%$ and $C=85\%$, which is achieved with $P_{mem}\geq0.75$. Results are in agreement with \cite{George2011, Castignani2016}.
    
    \item We analyze the distributions in different redshift ranges and masses and find a better selection of true galaxy members in group scales and for higher redshift ranges. This result is probably related to the cluster spatial distribution, where groups tend to be more concentrated.
    
    \item The comparison between redshift average values obtained through the selected member galaxies and the nominal ones, given by the mock, shows similar redshift values. This comparison also shows centre position variations around $64$ kpc between mock and selected galaxies, and $128$ kpc for PZWav and selected galaxies. 
\end{itemize}

We conclude that our estimator is robust to small offsets and produces optical mass proxies that are competitive with other traditional observables. 
Our method can be applied in present and future photometric surveys, such as S-PLUS and J-PAS, alongside any list of clusters or galaxy groups produced by cluster finders.

\section*{Acknowledgements}

LD was supported by a scholarship from the Brazilian federal funding agency \textit{Coordena\c{c}\~{a}o de Aperfei\c{c}oamento de Pessoal de N\'ivel Superior - Brasil} (CAPES).  ESC acknowledges the support of the funding agencies CNPq (\#309850/2021-5) and FAPESP (\#2023/02709-9). PAAL thanks the support of CNPq (grants 433938/2018-8 and 312460/2021-0) and FAPERJ (grant E-26/200.545/2023). R.A.D. acknowledges partial support from CNPq grant 312565/2022-4. The authors are thankful to the anonymous referee for the useful comments and suggestions.

\section*{Data Availability}
 
To access the mock catalogue presented in this article contact Pablo Araya-Araya (e-mail: paraya-araya@usp.br), asking for the specific dataset.  
The FAE and AME codes are available on GitHub: https://github.com/liadoubrawa/Probabilistic\_membership\_estimators.



\bibliographystyle{mnras}
\bibliography{AME_mass_proxies} 






\bsp	
\label{lastpage}
\end{document}